\documentclass[useAMS,usenatbib]{mnras}
\usepackage{graphicx,rotate,url,mathptmx,lscape,times}
\usepackage{./aas_macros}
\usepackage{color}
\title[SEDs and AGN Emission Lines]{State-of-the-art AGN SEDs for Photoionization Models: BLR Predictions Confront the Observations}

\author[G.J. Ferland, et al.]
       {\parbox[]{6.0in}
        {G. J. Ferland$^{1}$\thanks{E-mail: gary@uky.edu},
        C. Done$^2$,
        C. Jin$^{3,4}$,
        H. Landt$^2$,
        M. J. Ward$^2$\\
        \footnotesize
        $^1$Department of Physics, University of Kentucky, Lexington, KY 40506, USA\\
        $^2$Department of Physics, University of Durham, South Road, Durham DH1 3LE, UK\\
        $^{3}$National Astronomical Observatories, Chinese Academy of Sciences, 20A Datun Road, Beijing 100101, China\\
        $^{4}$School of Astronomy and Space Sciences, University of Chinese Academy of Sciences, 19A Yuquan Road, Beijing 100049, China\\
}}

\date{%Accepted .
      Received }

\pagerange{\pageref{firstpage}--\pageref{lastpage}}
\pubyear{}

\begin{document}

\maketitle

\label{firstpage}

\begin{abstract}

\noindent
The great power offered by photoionization models of Active Galactic Nuclei (AGN) 
emission line regions has long been mitigated by the fact that very little is known about the
spectral energy distribution (SED) between the Lyman limit, where intervening absorption
becomes a problem, and ~0.3 keV, where soft x-ray observations become possible.
The emission lines themselves can, to some degree, be used to probe the SED, but only in the broadest terms. This paper employs
a new generation of theoretical SEDs which are  internally self-consistent,
energy conserving, and  tested against observations, to infer properties of the emission-line regions. The SEDs are given as a function of the Eddington ratio, allowing emission-line correlations to be investigated on a fundamental basis.
We apply the simplest possible tests, based on the foundations of photoionization theory,
to investigate the implications for the geometry of the emission-line region.
The SEDs become more far-ultraviolet bright as the Eddington ratio increases, so the  equivalent widths
of recombination lines should also become larger,
an effect which we quantify.
The observed lack of correlation between Eddington ratio and equivalent width shows
that the cloud covering factor must decrease as Eddington ratio increases. This would be consistent with recent models proposing that the broad-line region is a failed dusty wind off the accretion disc. 

\end{abstract}


\section{Introduction}
\label{intro}

It has long been the goal of AGN astrophysics to be able to
use quasar emission lines to probe the centers of massive galaxies across the universe.
Photoionization models are often used to do this.  Reviews of such work
are given in the conference volume \citet{FerlandSavin01} and in the textbook
\citet{AGN3}, hereafter AGN3.

The emission lines are most sensitive to the spectral energy distribution (SED)
in the unobservable FUV, EUV, and XUV regions.
The interstellar medium blocks our view of this
spectral region, so indirect methods must be used to predict this part
of the radiation field.
The emission lines also carry information about this part of the SED, and the lines can be used
to constrain that region, as was done by \citet{MathewsFerland87}.
Most of the observed variation in emission line ratios can be characterised in a principle component analysis
\citep{1992ApJS...80..109B}. 
The dominant component
is Eigenvector 1, which most probably represents changes in the Eddington ratio:
$L/L_{\rm Edd}$ where $L$ is the bolometric luminosity and $L_{\rm Edd}$ is the Eddington luminosity, with smaller but still significant changes along Eigenvector 2, probably representing changes in 
mass \citep{1992ApJS...80..109B}.

A new generation of AGN SEDs have now become available, as summarized in
\citet{Done.C12Intrinsic-disc-emission-and-the-soft},
\citet{Jin.C12A-combined-Optical-and-X-ray-Spectra},
\citet{Jin.C12B-combined-optical-and-X-ray-study},
and \citet{Jin.C12AA-combined-optical-and-X-ray-study},
These papers model the AGN continua from nearby, well studied objects,
and stack them as a function of $L/L_{\rm Edd}$, giving a sequence of spectra which are ideal for pursuing such
questions as the origin of the Eigenvector emission line sequence
\citep{Boroson.T92The-emission-line-properties-of-low-redshift-quasi-stellar,
2010MNRAS.409.1033M}. The aim of this paper is to test whether the expected changes 
in the SEDs 
are consistent with the observed changes in emission line properties across this sample
(see also \citet{2019ApJ...875..133P} for a similar study using H$\beta$ and FeII).
There have been a number of studies which have tried to derive the SED directly from 
observations, starting with \citet{MathewsFerland87} and recently by 
\citet{ 2011ApJ...738....6M}.  These studies are entirely complementary to the current approach,
which begins with ad initio self-consistent models of the SED.
One of our goals is to propose simple but robust tests for the validity of these new models.

The optical line emission can be predicted using large photoionization codes such as Cloudy 
\citep{CloudyReview13}. However, the differences can be illustrated more powerfully using 
simple first principles of photo-ionization as these give constraints simply from photon number counts. 
The set of SEDs used here have harder FUV spectra as $L/L_{\rm Edd}$ increases so we probe this using the broad line from He II 
(4686\AA ) as this is produced by recombination from completely ionized He, requiring photons above 52~eV. 
However, line equivalent widths can be affected by reddening as well as systematic uncertainties on geometry 
(covering factor, continuum anisotropy etc.). Hence we also compare He II with
broad H$\beta$ as this is nearby at 4861\AA~so is similarly affected by reddening but is from hydrogen recombination so requires
photons above 13.6~eV. The SEDs predict that both lines should increase in EW with $L/L_{\rm Edd}$. 
This is not 
seen in the data so it requires a systematic decrease in BLR emissivity to compensate for the predicted change,
as shown in the analysis below (e.g. sec 3). 
The models correctly predict that the ratio of HeII/H$\beta$ should increase by a factor $\sim 2$ from the harder FUV spectrum as $L/L_{\rm Edd}$ increases from $0.1\to 1$, and by another factor of 2 
for the most extreme super-Eddington Narrow Line Seyfert 1 (NLSY 1) known. The data are compatible with this, though the
models form an upper envelop to the observations. Either the SED change seen in the data is not seen by the BLR
(anisotropic and/or filtered emission) or the 
covering fraction of the high ionization lines changes by more than that of the low ionization lines.

\section{Overview of the SEDs}

The SEDs used here are based on the study of \citet{Jin.C12AA-combined-optical-and-X-ray-study}. This uses a 
sample of unobscured AGN with both SDSS spectra covering broad H$\beta$ and good signal-to-noise {\it XMM-Newton} UV/X-ray observations,
giving a well sampled SED together with black hole mass estimator (see also \citealt{Jin.C12A-combined-Optical-and-X-ray-Spectra}). These 51 AGN were fit using the SED models
developed by \citep{Done.C12Intrinsic-disc-emission-and-the-soft}. 
It has long been known that the observed broad-band SEDs of AGN are not well reproduced by a standard Shakura-Sunyaev disk 
with ``extra'' emission in the IR and X-ray \citep{Elvis1994}. The UV turns down at energies below the expected peak for a disk which extends down to the last stable circular orbit, there is an X-ray tail which extends out to high energies, and there is a soft X-ray upturn which appears to point upwards to match the UV downturn (soft X-ray excess component). Nonetheless, the 
optical/soft UV spectrum is generally quite well matched by the  disc emission expected from a 
standard accretion disc. 

In the standard disk models, the mass accretion rate is constant with radius, so the mass accretion rate through the outer disc sets the mass accretion rate through the entire accretion flow, irrespective of its structure. Assuming that the emissivity is standard Novikov-Thorne sets the luminosity emitted at each radius, 
and the full SED can then be fit assuming that this energy thermalises at large radii, 
but below some coronal radius, $R_{cor}$, determined by fitting the data, 
the energy is instead emitted as a combination of warm, optically thick Comptonisation (to fit the UV downturn and soft X-ray excess) or hot, optically thin Comptonisation (to fit the high energy X-ray tail). This gives a model which has enough components to follow the data, but with physical limitations on the energetics which allows the fits to be well constrained. 

The models have undergone some development since the first study of \citet{Jin.C12A-combined-Optical-and-X-ray-Spectra}.
This first paper assumed that the outer disk emission thermalised to a blackbody at the effective temperature ({\sc optxagn} model in {\sc xspec}), whereas electron scattering within the disc atmosphere is expected to cause a shift in the observed temperature once hydrogen becomes ionised, giving a colour temperature correction (referred to as $f$) to the emission ({\sc optxagnf} model in {\sc xspec}: \citep{Done.C12Intrinsic-disc-emission-and-the-soft}). This latter model is used in \citet{Jin.C12AA-combined-optical-and-X-ray-study}, and we base our study here on the results from this. A subsequent model upgrade which treats the soft Compton component more exactly was developed by \citet{Kubota18}, but these have yet to be fit to the data. 

\citet{Jin.C12AA-combined-optical-and-X-ray-study} split the sample into three sub-samples in every parameter to try to determine which was the most likely to be responsible for changes in the observed SED. To some extent, looking for a single parameter family should be doomed to failure as there is considerable spread in both mass and mass accretion rate across the sample, let alone additional scatter which could be introduced from a range in black hole spin and/or inclination angle (though hopefully the latter is small as significantly obscured objects are not included). They concluded that most of the variance was due to $L/L_{\rm Edd}$.
The three SEDs resulting from the stacked low, medium and high $L/L_{\rm Edd}$ sub-samples have mean 
$\log L/L_{Edd}=-1.15,-0.55, -0.03$, respectively, and are show in Fig.\ref{fig:SED} normalised to the wavelength of H$\beta$. The major change 
in shape is the increase in disc emission in the UV, correlated with a softer X-ray tail. In the model, this implies a decreasing radius, $R_{cor}$ at which the standard disc transitions to the Comptonised components (decreasing from $155\to 60\to 13$), and to a softer spectral index of the high energy tail (photon index $\Gamma$ increasing from $1.8\to 1.9\to 2$).
For comparison, the SED deduced by  \citet{MathewsFerland87}
is shown as the dotted line.

We extend this set of SEDs to even higher $L/L_{\rm Edd}$ by including the single object RX J0439.6-5311 (\citealt{2017MNRAS.471..706J}). This is one of the most extreme Narrow Line Seyfert 1 galaxies (NLS1), and has an exquisitely well determined SED as there is very little interstellar extinction along this line of sight, so the continuum can be seen directly to 912\AA~in the rest frame ({\it HST} COS), and is visible again at 0.1~keV ({\it ROSAT}). Again, normalising at H$\beta$ shows that this continues the trend in SED properties, being even more dominated by the disc component, with even steeper X-ray tail. We use this set of 4 SEDs spanning nearly two orders of magnitude in $L/L_{\rm Edd}$
to calculate the expected change in H$\beta$ and He II line emission. The models clearly show differential change in the 52~eV continuum relative to the 13.6~eV continuum (marked by vertical lines) so there should also be changes in the He II and H$\beta$ line emission.

\section{The ``Zanstra temperature'' test of recombination line equivalent widths}
This section applies the test first suggested by \citet{1929PDAO....4..209Z} and described in
AGN3 Section 5.10.
The idea is simple.  The luminosity of an H or He recombination line is proportional
to the number of ionizing photons in the H or He-ionizing continuum.
Each ionizing photon produces one photoionization, resulting in one recombination, so the
lines count the number of photons in the FUV-EUV-XUV part of the SED.  
The equivalent width of an H or He recombination line is proportional
to the ratio of the number of ionizing photons 
to the continuum under the emission line, so is
a direct probe of the SED shape.
While many other UV and optical lines will change too, they are far more difficult to 
model so are left for a later paper.

We first show the idea in terms of simple recombination theory.
Detailed calculations show that each hydrogen-ionizing photon produces one hydrogen ionization 
(AGN3).
If the central object is surrounded by clouds that are optically thick in the Lyman continuum,
then the number of ionizing photons absorbed by clouds per second will be
\begin{equation}
N_{ion} =  \frac{\Omega}{4\pi}  Q(H)\ \ \ [s^{-1}]
\label{eq:QH}
\end{equation}
where Q(H) is the total number of ionizing photons
in the SED [s$^{-1}$, AGN3 Sec 2.1] 
and $\Omega / 4\pi$ is the gas covering factor, the fraction of the ionizing
photons which strike gas and are absorbed.
Photoionizations and recombinations are in balance, so Equation \ref{eq:QH} is also 
the number of hydrogen recombinations per second.
The ratio 
$\alpha_{eff}(H\beta) / \alpha_B(H) \sim 1/8$ is the fraction of hydrogen recombinations
which produce an H$\beta$ photon in Case B conditions (AGN3 Section 4.2).  
The number of H$\beta$ photons emitted per second is then
\begin{equation}
N(H\beta, Case\ B) =  Q(H) \frac{\Omega}{4\pi}   \frac{\alpha_{eff}(H\beta)}{\alpha_B(H)}  \ \ \ [s^{-1}]
\label{eq:NHb}
\end{equation}
The equivalent width of the line, $EW(line)$, is then
\begin{equation}
EW(H\beta, Case\ B) = h\nu(H\beta)  \  \lambda(H\beta) \ \frac{\Omega}{4\pi} 
\frac{\alpha_{eff}(H\beta)}{\alpha_B(H)} \frac{Q(H)}{ \nu F_{\nu}(H\beta)} 
\label{eq:Zanstra}
\end{equation}
which is approximately given by
\begin{equation}
EW(H\beta, Case\ B) \approx 4.81\times 10^{-13} t_4^{-0.06}  \frac{\Omega}{4\pi} 
\frac{  \lambda(H\beta) Q(H)}{ \nu F_{\nu}(H\beta)} 
\end{equation}
where $t_4$ is the gas temperature in units of 10$^4$~K and the recombination coefficients given in
AGN3 are used.
The equivalent equation for He II $\lambda 4686$ is
\begin{equation}
EW(\lambda 4686, Case\ B) \approx 9.35\times 10^{-13} t_4^{-0.28}  \frac{\Omega}{4\pi} 
\frac{  \lambda(4686) Q(He^+)}{ \nu F_{\nu}(\lambda 4686)} 
\label{eq:ZanstraHeII}
\end{equation}
In this way the equivalent width of a recombination line such as H$\beta$ 
or He II $\lambda 4686$ is determined by the 
SED and gas covering factor.
This is the ``Zanstra method'' of determining stellar temperatures 
(see AGN3, Sec 5.7).  
Although the ideas presented here are based on foundation recombination theory,
they are in accord with the BLR model predictions presented below.

The equivalent width $EW(H\beta)$  
depends on the cloud covering factor $\Omega/4\pi$ since only 
$\Omega/4\pi$  of  ionizing photons strike clouds,
are absorbed, and produce an H$\beta$.  
If all AGN have  the same $\Omega/4\pi$  then $EW(H\beta)$
directly probes the SED hardness of various objects.
Alternatively, with models of the SED shape such as those described here,
the equivalent width can be used to determine whether the covering factor
depends on other parameters of the black hole.

\begin{figure}
\begin{center}
\includegraphics[clip=on,width=\columnwidth,height=0.8
\textheight,keepaspectratio]{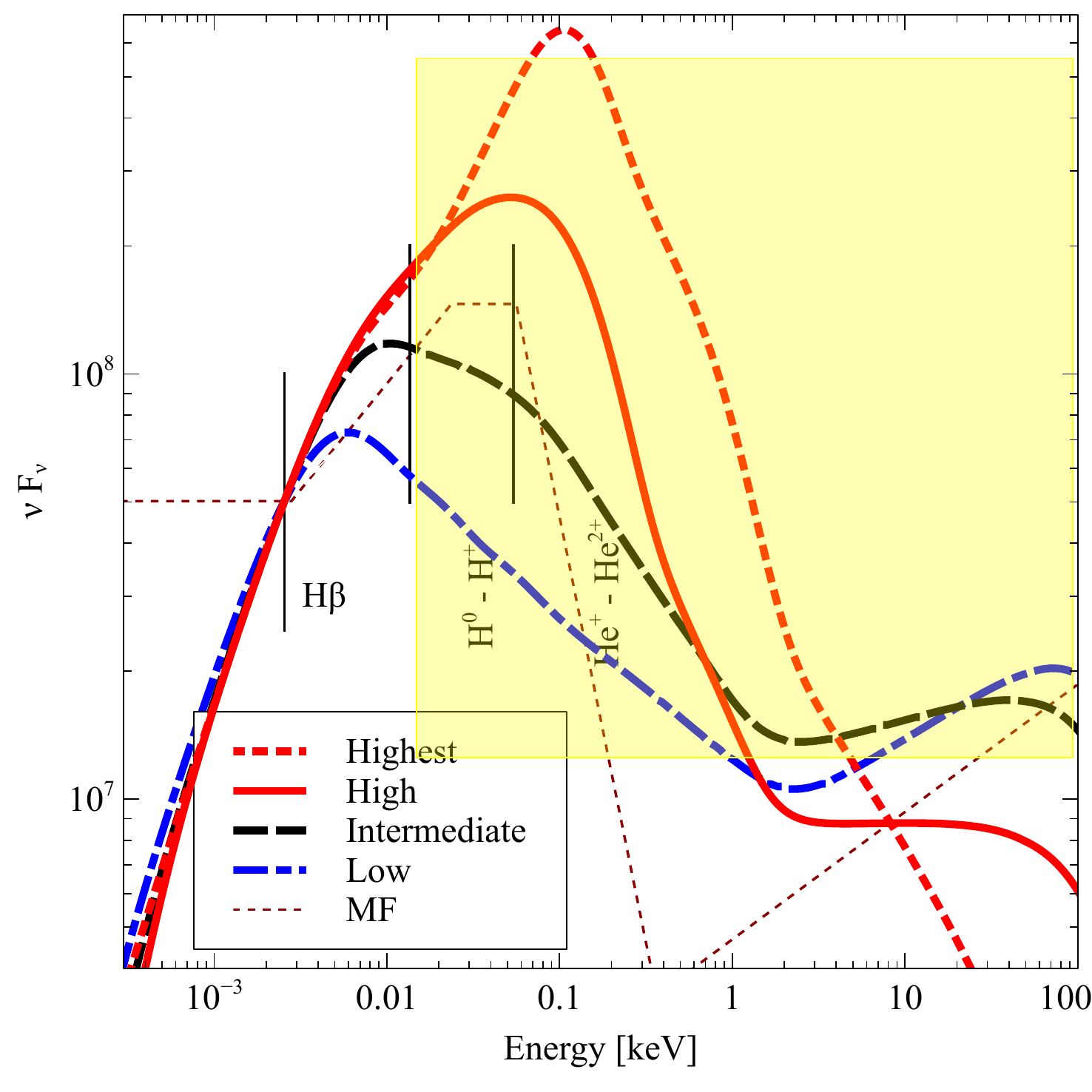}
\end{center}
\caption{The four SEDs studied in this paper are compared.  They are
normalized to have the same intensity at the wavelength of H$\beta$,
to facilitate comparison with line equivalent widths.
The yellow region marks the hydrogen-ionizing part of the SED.
The equivalent width of a recombination line such as H~I H$\beta$
is proportional to the ionizing photon luminosity within the yellow region.
The energy to fully ionize He and produce He II emission is also marked.
The dotted line marked ``MF'' is the mean SED deduced by  \citet{MathewsFerland87}. }
\label{fig:SED}
\end{figure}

Figure \ref{fig:SED} compares the four SEDs discussed above.
The photon energy is given in keV, the units used in the original SED papers, and  
the  SEDs are normalized to have the same intensity at the wavelength of H$\beta$. 
This makes it simple to compare optical recombination line equivalent widths.
 The last term in Equation \ref{eq:Zanstra} is the ratio of the area in yellow 
in Figure \ref{fig:SED} to the SED intensity at the wavelength of H$\beta$.
The equivalent width is proportional to the
number of photons continued in the yellow regions
of the SEDs.
The figure also marks the ionization limit for fully ionizing He, 54.4 eV.
This part of the SED controls He II emission.

The number of ionizing photons is predicted to
strongly correlate  with $L/L_{\rm Edd}$.
The ``Highest'' SED produces  the most Lyman continuum photons relative to the continuum at 
H$\beta$ so its H$\beta$ line will also have the highest equivalent width,  $\sim 1$ dex
higher than the lowest Eddington ratio SED.
The number of photons that can ionize helium increases even more.
These considerations provide a strong, model independent, prediction that the equivalent widths
of H I and He II will
strongly correlate with the Eddington ratio.
Although Figure 1 shows that a significant amout of energy is emitted at high energies,
the tests proposed here are based on photon counting and there are relatively few high-energy photons.
The SED around the ionization potentials of H and He has the most important effect on our tests.

Table \ref{tab:EWHb} compares the predicted H$\beta$ and He~II $\lambda 4686$ equivalent widths for the four SEDs.
The rows marked ``H$\beta$ Q(H)'' and ``He II Q(He$^+$)'' are straightforward applications of Equations \ref{eq:Zanstra} and \ref{eq:ZanstraHeII} assuming Case B.
This should be quite accurate for lower density gas, for instance,  the NLR.
The situation for H I in the BLR is more complex due to the high 
resulting line optical depths, Ly$\alpha$ trapping, and the
importance of photoionization from excited states \citep{1990agn..conf...57N}
but these processes have a much smaller effect on He II since its resonance lines are
destroyed by photoionization of hydrogen \citep{1985ApJ...299..785E} with the 
result that He II should be closer to Case B.
The  rows marked ``BLR'' give predictions for a ``standard'' BLR cloud 
(log U =-1, log N(H) = 10) using Cloudy \citep{2017RMxAA..53..385F}.
The H I equivalent widths are generally within $\sim 50\%$ of Case B while He II is
even closer to Case B.  The strong trend for  increasing Eddington
ratio to produce larger equivalents is obvious.
The highest-Eddington-ratio objects should have recombination line equivalent
widths roughly 1 dex larger than the low-Eddington-ratio objects.
For reference, the ratios of changes in  the ratio of number of ionizing photons for 
H and He and the ratio of ionizing to non-ionizing photons as a function of Eddington ratio
is given by the ratios of predicted equivalent widths in this Table.

% excel He2_EQ, BLR_runs in SED_draft / 1906figs, 
% BLR calculations are in 
% /Users/gary/Dropbox/papers/Ward_SEDs/figs/BLR_meanSEDs\ 19\ 06\ 17\ f92                                               

\begin{table}
\caption{Predicted and observed equivalent widths for H 1 $\lambda 4861$ and He II $\lambda 4686$.
}
\begin{center}
\begin{tabular}{@{}ccccc@{}}
\hline
$EW$(Line) & Low & Inter & High & Highest\\
\hline
H$\beta$ Q(H) &90.1\AA & 198\AA & 432\AA & 643\AA \\
H$\beta$ BLR &  128\AA  & 268\AA  &  527\AA & 828\AA   \\
H$\beta$ Obs &  68 $\pm$34\AA & 93 $\pm$33\AA  &  76 $\pm$39\AA & 25.6 $\pm$1.1\AA  \\
\hline
He~II Q(He$^+$)  & 26.0\AA & 63.0\AA & 181\AA & 492\AA   \\
He II BLR       & 21.2\AA & 50.7\AA & 145\AA & 403\AA  \\
He II Obs        & 12.1$\pm$9.9\AA & 16.4$\pm$11.2\AA & 10.9$\pm$8.1\AA & 2.1$^{+3.6}_{-2.1}$\AA  \\
\hline
$\Omega/4\pi$(H$\beta$) & 0.755$\pm 0.377$ & 0.470$\pm 0.167$ & 0.176$\pm 0.090$ & 0.040$\pm 0.002$ \\
$\Omega/4\pi$(He II) & 0.465$\pm$0.381 & 0.260$\pm$0.178 & 0.060$\pm$0.045 & 0.0043$^{0.0073}_{0.0043}$ \\
\hline
\end{tabular}
\end{center}
\label{tab:EWHb}
\end{table}

It is likely that clouds actually have a distribution of parameters, the Locally Optimal-Emitting Cloud (LOC) model of the emitting regions \citep{Baldwin1995}.
Calculations show that the equivalent widths of the recombination lines
we use here are consistent with large regions of BLR parameter space
\citep{KoristaBaldwin1997} for a particular SED.
Our goal here is to make differential comparisons over the range of Eddington ratio,
to document the effects of changes in the SED.
Whatever differences are introduced by the LOC model should not
affect our differential comparison \emph{if the structure of the BLR does not change
as the Eddington ratio changes}.
The following sections argue that large structural changes are, in fact, taking place.

\section{Comparison with observations}
\begin{figure}
\begin{tabular}{c}
\includegraphics[trim=0.9in 2.8in 0.0in 3.6in, clip=1,scale=0.48]{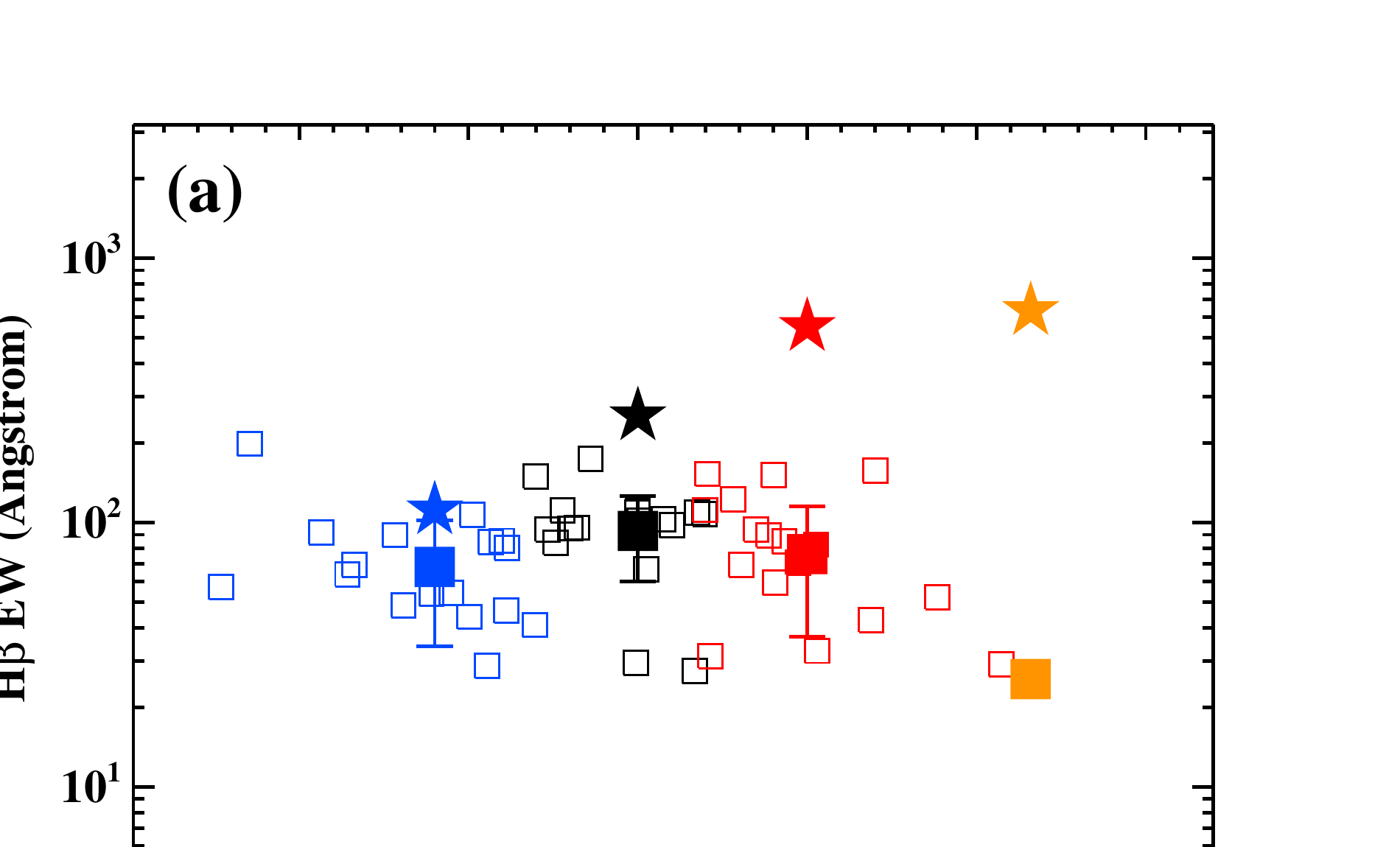}\\
\includegraphics[trim=0.9in 2.8in 0.0in 3in, clip=1,scale=0.48]{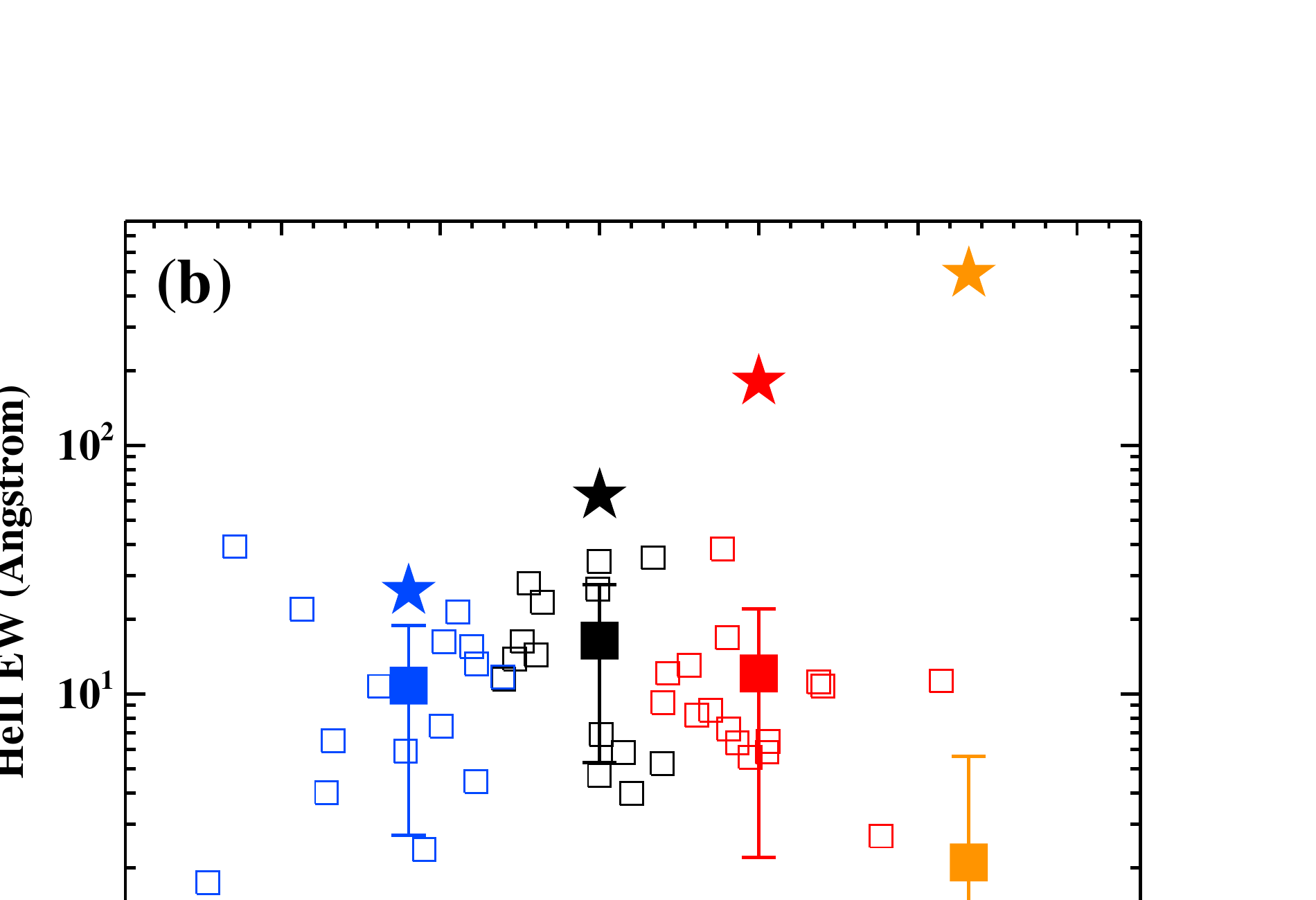}\\
\includegraphics[trim=0.9in 1.8in 0.0in 3in, clip=1,scale=0.48]{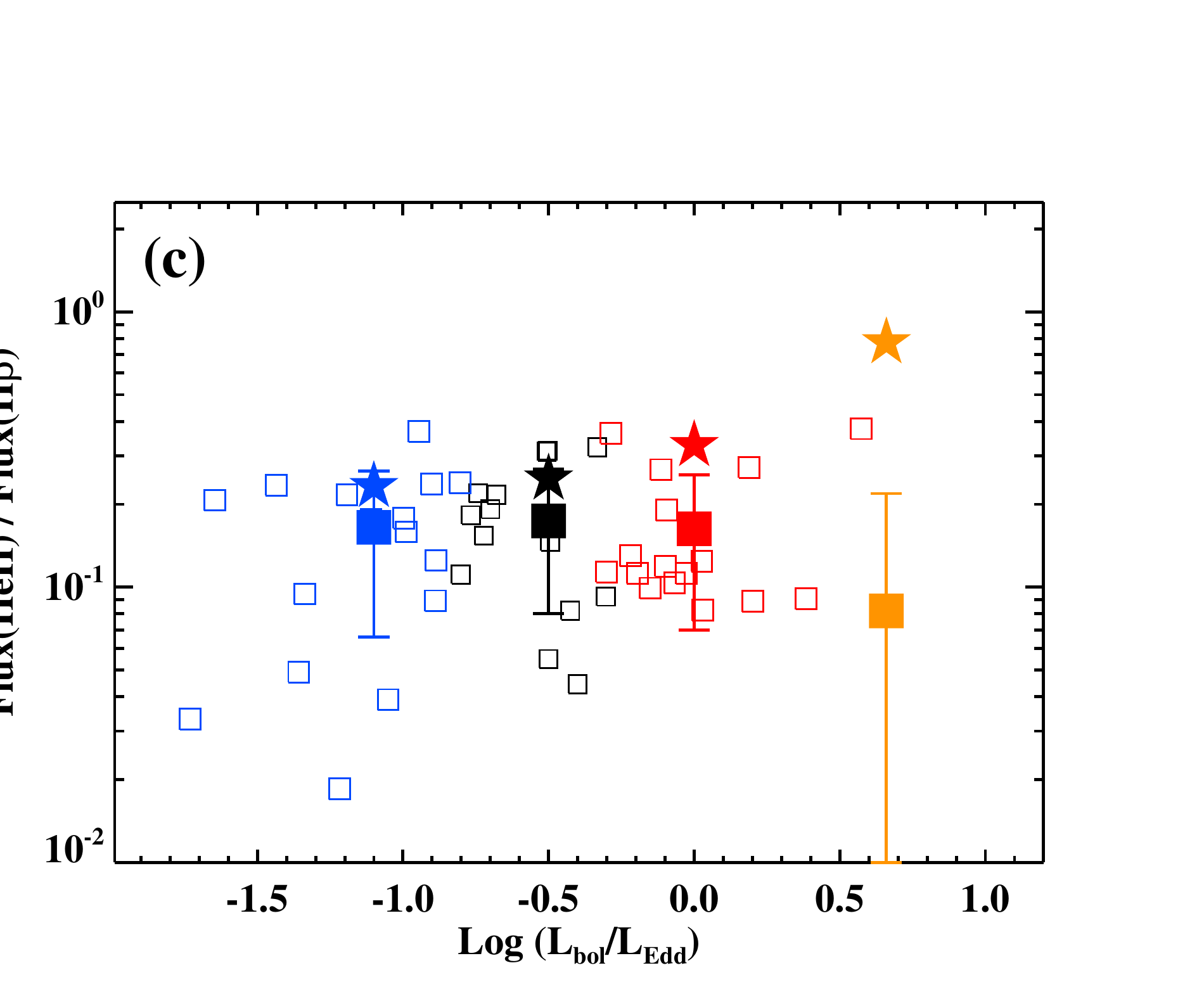}\\
\end{tabular}
\caption{The  H$\beta$ (top) and He~II $\lambda 4686$ (middle) equivalent widths as a function of the Eddington ratio. The filled squares represent the observed values for the Eddington-ratio groups. The coloured stars give the predicted equivalent widths assuming that clouds fully cover the continuum source. Larger EW's occur for higher Eddington ratio SEDs because of relatively more FUV flux. These predictions scale linearly with the cloud covering factor. The lower panel gives the ratio of the line equivalent widths (or fluxes), for He II / H I. This has the advantage that the cloud covering factor cancels out, and so it should have smaller cosmic scatter than the upper two panels.}
\label{fig:He2Hb}
\end{figure}

The mean  equivalent widths for the four Eddington ratio groups,
as measured from the sample described by
\citet{Jin.C12A-combined-Optical-and-X-ray-Spectra} and
\citet{2017MNRAS.471..706J}, 
are listed in the ``obs'' rows
of Table  \ref{tab:EWHb}.
The uncertainties in the first three groups represents the ensemble average while
the highest $L/L_{\rm Edd}$ gives the measurement uncertainty in the single object.
The equivalent widths for individual objects in the sample are shown in Figure \ref{fig:He2Hb}, along with the mean,
with different plot symbols used to indicate the  sub-classes.

Two tests can  be performed.
The first, shown in the upper and middle panels of Figure \ref{fig:He2Hb},
is the line equivalent width as given by
Equations \ref{eq:Zanstra} and \ref{eq:ZanstraHeII}.
The large stars give the predicted equivalent widths for the four
groups in Table  \ref{tab:EWHb}.
These assume that clouds fully cover the continuum source,
that is, $\Omega / 4 \pi = 1$ in 
Equations \ref{eq:Zanstra} and \ref{eq:ZanstraHeII}.
The prediction that the equivalent width should increase as the Eddington ratio
increases and the SED at the ionization energies becomes harder is clear.

%
% speadsheets in 1906figs
% upper left
% upper right  He2_EQ sent
% lower
%
%\begin{figure*}
%\begin{center}
%\includegraphics[clip=on,width=16cm,height=1.2
%\textheight,keepaspectratio]{3way.pdf}
%\end{center}
%\caption{The  H$\beta$ (left) and He~II $\lambda 4686$ (right) equivalent widths
%as a function of the Eddington ratio.
%The filled squares represent the observed values for the 
%Eddington-ratio groups.
%The colored stars give the predicted equivalent widths assuming that clouds
%fully cover the continuum source. 
%Larger EW's occur for larger Eddington ratio because the SED is harder.
%These predictions scale linearly with the cloud covering factor.
%The lower panel gives the ratio of the line equivalent widths
%or fluxes, He II / H I.
%This has the advantage that the cloud covering factor cancels out,
%so should have smaller cosmic scatter than the upper two panels.
%}
%\label{fig:He2Hb}
%\end{figure*}

The observations have a large scatter but do not follow the expected changes in equivalent width.
The simple expectations of Equation \ref{eq:Zanstra} 
and  \ref{eq:ZanstraHeII} do not agree with observations.
This could mean that the SED is incorrect, or that the cloud geometry depends on the Eddington ratio.
Theory and observation can be reconciled by setting the covering
factor to the ratio of 
the observed to theoretical equivalent widths.  
These required covering factors, given as the ratio of the observed equivalent width to ``Q(H)'' and
``Q(He$^+$) predictions, are given in the last two
rows of Table \ref{tab:EWHb}.

In the next section we make a qualitative suggestion for how this might occur.

\section{Discussion and conclusions}

One possibility for the lack of correlation between $L/L_{\rm Edd}$ and equivalent width
is that the clouds do not ``see'' the same SED that we do \citep{1997ApJ...487..555K}.
Alternatively, the geometry of the BLR could be a function of the SED, an idea that we will now discuss. 

The cloud covering factor given in Table \ref{tab:EWHb} decreases as the
Eddington ratio increases.  
The consequences of changing covering factors have been discussed by,
among many others, \citet{2015ApJS..219....1O},
\citet{2019ApJ...870...26L} and
\citet{2013ApJ...777...86L}.
The low Eddington ratio objects  require 
covering factors around 50\%,
with the covering factor decreasing to $\sim 5\%$ for the highest Eddington ratio case.
Such changes are not totally ad hoc.
It has previously been suggested that an inverse correlation between covering factor and
luminosity is the explanation for the ``Baldwin effect'', the tendency for the UV C~IV 
line equivalent width to decrease with increasing luminosity
\citep{1984ApJ...278..558M}.
X-ray observations further suggest that the covering factor
varies over the range suggested by Figure \ref{fig:He2Hb}
and becomes larger as the luminosity decreases 
\citep{2013A&A...553A..29R,2013arXiv1303.0219L}.

The ratio of the line equivalent widths, EW(He~II~4686) / EW(H$\beta$),
measures the number of photons with $h\nu > 54$~eV relative to the number
with $h\nu > 13.6$~eV.
This ratio has the advantage that the cloud covering
factor cancels out,  so it might be expected to have a
smaller scatter than the equivalent widths.
This ratio is shown in the lowest panel of Figure \ref{fig:He2Hb}.
Curiously, the predicted values hover over the upper envelope of the scatter.
The difference in the covering factors of the H I and He II
lines is within the uncertainties of the approximations 
made in our analysis.
There is a hint of a trend in the data which parallels the theoretical expectation
given in Equations 1-5.
This supports the suggestion that decreasing covering factor could account for the 
discrepancy seen in the upper two panels. Now we consider the lower panel which traces the line ratio. 
The track of the predicted line ratio lies consistently above the observations, and particularly so for the highest SED.
Another way of putting this is that the high-ionization lines change by more than the low ionization,
requiring that their covering factor also change more.

The factor of 10 difference between the observed and theoretical H$\beta$ and He II lines strengths
 for the high Eddington ratio objects is a due to the change in the ionizing continuum with Eddington ratio 
 in our adopted model. 
 It has long been known that the ratio of the x-ray to UV is a function of luminosity 
 (see \cite{2018ApJ...860...41L} for the latest result), 
 and a function of Eddington ratio but with very large scatter \citep{2010A&A...512A..34L}. 
 We know of no systematic study of the effects on the line spectrum of these  changes in the SED.
 The existence of a set of comprehensive and self consistent SEDs makes this now possible.
 This paper is a first step in that direction.

The BLR is an optically thick structure illuminated anisotropically by
the accretion disk flux. As the LOC model of the BLR shows, for a given
range in number density, each emission line radiates most efficiently
at a given ionising {\it flux} \citep[e.g.][]{2004ApJ...606..749K}. 
The location where this (constant)
flux is reached will depend on the (ionizing) luminosity (of which
$L/L_{\rm Edd}$ is a good measure); it will increase with increasing
luminosity (as $R \propto \sqrt{L}$). In this scenario, our
observation of a decreasing BLR covering factor with increasing
Eddington ratio can be explained if the BLR structure had a constant scale
height. Then, as the LOC radius increases, its half-opening angle
$\Phi$ (related to the covering fraction CF as $CF=sin(\Phi)$) as seen
by the central AGN decreases, resulting in a reduced EW. A BLR with a constant scale height located in the equatorial plane is expected under the recent models of 
\citet{2011A&A...525L...8C, Czerny17} and \citet{Baskin18}, in which the BLR is a failed dusty wind off the accretion disk. 

For the super-Eddington AGN, the change of disc structure may become important as well, which can further modulate the intensity of different emission lines. In this extreme case, the presence of a puffed up inner disk region, and/or a clumpy strong disk wind, may partially shield the ionizing flux originated from the high energy band of the SED including the soft X-ray excess and part of the UV emission. This would affect He II more than H I because the ionizing energy of He II is higher (see Figure \ref{fig:SED}), and so reduce the ratio between these two lines. This concept is discussed for super-Eddington AGN and shown as a cartoon in \citet{2017MNRAS.471..706J}.

It has become common to estimate AGN black hole masses from
single-epoch spectra using the emission line luminosity as a proxy for
the BLR radius instead of the optical continuum luminosity. This is
because the continuum emission can suffer from contamination by the host
galaxy flux in low-z sources, or by non-thermal jet emission in
radio-loud sources. Our investigation predicts that for high
$L/L_{\rm Edd}$ sources, the use of the line luminosity will under predict
the black hole mass by factors of several because of the required changes
in cloud covering factor.

Our results also suggest that changes in the SED alone 
are not responsible for emission line differences between NLS1,
which are thought to have high Eddington ratios and so represented by
our ``highest'' case, and 
broad line Seyferts. We propose that the geometry of the BLR is also likely to change.

The next step would be to study a much larger AGN sample that can be divided according to $L/L_{\rm Edd}$, black hole mass and spin.

\section{Acknowledgments}
GJF acknowledges support by NSF (1816537), NASA (ATP 17-ATP17-0141), and STScI (HST-AR- 15018). C.J. acknowledges the National Natural Science Foundation of China through grant 11873054, as well as the support by the Strategic Pioneer Program on Space Science, Chinese Academy of Sciences through grant XDA15052100. C.D., H.L. and M.J.W. acknowledge the Science and Technology Facilities Council (STFC) through grant ST/P000541/1 for support.

\bibliographystyle{mnras}
\bibliography{SEDs}
\bsp

\label{lastpage}
\clearpage
\end{document}